\documentclass[showpacs,preprintnumbers,amsmath,amssymb,twocolumn]{revtex4}

\usepackage{graphicx}
\usepackage{dcolumn}
\usepackage{bm}
\usepackage{color}

\begin{document}


\title{Noether gauge symmetry of Dirac field in 2 + 1 dimensional gravity}

\author{Ganim Gecim}%
   \email{ganimgecim@akdeniz.edu.tr}
\author{Yusuf Kucukakca}%
   \email{ykucukakca@akdeniz.edu.tr}
   \author{Yusuf Sucu}%
   \email{ysucu@akdeniz.edu.tr }

 \affiliation{Department of Physics,
Akdeniz University, Antalya, Turkey
   }%
\date{\today}
\begin{abstract}
In this paper, we consider a gravitational theory
including a Dirac field that is non-minimal coupled to gravity in
$2+1$ dimensions. Noether gauge symmetry approach can be used to fix
the form of coupling function $F(\Psi)$ and the potential $V(\Psi)$
of the Dirac field and to obtain a constant of motion for the
dynamical equations. In the context of $2+1$ dimensions gravity, we
investigate cosmological solutions of the field equations using
these forms obtained by the existent of Noether gauge symmetry. In
this picture, it is shown that for the non-minimal coupling case,
the cosmological solutions indicate both an early-time inflation and
late-time acceleration for the universe.
\end{abstract}

\maketitle

\section{\label{sec:intro}Introduction}
The general theory of gravitation describes the physical phenomena
coming about in a strong gravitational field and the physical
behaviour of the Universe in a large-scale. At the same time, it has
a remarkable mathematical and physical perspective. However, as our
knowledge about the universe increases by new observational data,
some new modified gravitation theories based on the theory are
appeared to understand the reality of the universe. Moreover, the
quantisation of the theory still keeps standing the most important
problem in physics. Therefore, the general theory of gravitation is
one of the most active research areas.

Giving the observational data \cite{ade1,sper1,sper2}, the
universe had been accelerated in early time which is called
inflation. To investigate the cosmic inflation of the universe or
the beginning of expansion, the standard cosmological models were
used \cite{alan,linde}. Another cosmic acceleration of the universe
occurs in the late-time universe and confirms various observational
evidences which are the observations of supernovae Type Ia  (SNe Ia)
\cite{riess99,Perll}, cosmic microwave background radiation (CMB)
\cite{netterfield02,bennet}, and large-scale structure
\cite{tegmark04}.  To clarify such an accelerated expansion, many
authors introduced mysterious cosmic fluid, the so-called dark
energy. It has been proposed severel models in the literature such
as quintessence  \cite{Rat}, phantom \cite{Cad}, $F(R)$
\cite{Sot} and $F(T)$ gravity \cite{Beng}.
Recently, to understand the problem in 3+1 dimensional spacetime, it
has been realized that fermionic or Dirac fields as a gravitational
source which cause inflationary period in early universe and dark
energy in old universe are being considered
\cite{Saha1,Ar,Saha2,Rib1,Rib2,vakili,Kremer2}. In this connection,
Sousa and Kremer \cite{Kremer} have analysed a model with Dirac
field that is non-minimally coupled to the gravity in the $3+1$
dimensional. They have utilized Noether symmetry approach to
determine the forms of the potential and the coupling function to
show dependence on Dirac fields in the model. Noether symmetry
approach firstly introduced by Cappoziello et al. \cite{capo1} to
find new cosmological solutions in the $3+1$ dimensional gravity has
been used to determine the shape of the potential and the coupling
function dynamically in the scalar-tensor gravity theory. It is
important to note that this approach gives us constants of motion
(first integral) for the dynamical equation. Recently, this approach
has been extensively studied in various cosmological models i.e., in
scalar field cosmology \cite{capo2}, $f(R)$
metric and Palatini theory \cite{cap03}, $f(T)$
theory \cite{wei}, telleparalel dark energy model \cite{kuc}.
All of these studies, it has been considered the Noether symmetry
approach without a gauge term. If one consider  the gauge term, then
he may expect extra symmetries which yield a new constants of
motion, but the term makes the calculations more complicating.

The general theory of gravitation in $3+1$ dimensions is very
difficult because it has complicated calculations. However, in $2+1$
dimensional spacetime, it takes much more simple form because Weyl
tensor vanishes and Riemann tensor is reduced to Ricci tensor.
Moreover, the version in the $2+1$ dimensions also contains the
physical results in the $3+1$ dimensional theory
\cite{wit,deser,cor,banados}. Therefore, the $2+1$ dimensional
spacetime becomes a perfect theoretical laboratory to construct new
modified gravitation theories. In particular, if one wants to probe
the mathematical and physical properties of the universe in the
$2+1$ dimensional spacetime by using Dirac spinor fields as a
gravitational source, one can simply describe the properties of the
universe according to its analogous $3+1$ dimensional spacetime
because Dirac spinor fields with 4-components in the $3+1$
dimensional spacetime are also reduced to the 2-components spinor
fields corresponding one negative and one positive energy states
\cite{sucu}. Furthermore, Dirac spinorial fields can give us useful
information about inflation in the early time of the universe
because Dirac theory has vacuum which includes zitterbewegung
oscillations between positive and negative energy states. Also, the
theory perfectly describe an interaction between Dirac particles and
matter. Therefore, Dirac spinorial fields can probe how to expand
the universe in late time.

From these points of view, we want to study the Dirac fields as a
source of early-time inflation and late-time acceleration in $2+1$
dimensional gravity for Friedmann-Robertson-Walker (FRW) background
by using Noether gauge symmetry approach. The results to be
performed in this study are important in consequence of the $2+1$
dimensional gravity and Dirac theory, because they give information
about the expansion of the universe in early and late time. Also, it
will be first example carried out, in regarding gravity and Dirac
fields by using Noether gauge symmetry method. Therefore, the study
will satisfy a motivations for new studies in the $2+1$
dimensional gravity.

This paper is organized as follows. In the following section, we
give the field equations of a theory in which the Dirac field is
non-minimal coupled to the gravity in the $2+1$ dimensions. In
Section \ref{ns}, we search the Noether gauge symmetry for the
Lagrangian of the theory with the Dirac field. In Section
\ref{fieldeqns}, we obtain the solutions of the field equations by
using Noether gauge symmetry approach. Finally, in the Section
\ref{conc}, we conclude with a summary of the obtained results.
Throughout the paper, we use $c=G=\hbar =1$.

\section{The action and field equations} \label{b1}

In $2+1$ dimensional curved spacetime, the action for a Dirac field
which is non-minimally coupled to scalar curvature, is given by
\begin{eqnarray}
\mathcal{A} =\int d^{3}x\sqrt{-g}\Bigg\{F(\Psi )R+\frac{\imath
}{2}\Big[( \bar{\psi}\bar{\sigma}^{\mu }(x)\left( \partial _{\mu
}-\Omega _{\mu }(x)\right) \psi \nonumber \\
-\bar{\psi}(\overleftarrow{\partial
_{\mu }}+\Omega _{\mu }(x))\bar{\sigma} ^{\mu }(x)\psi \Big]-V(\Psi
)\Bigg\},\raggedright \label{act1}
\end{eqnarray}
where $F(\Psi )$ and $V(\Psi )$ generic functions, representing the
coupling with gravity and the self-interaction potential of the
Dirac field respectively, and they depend on only functions of the
bilinear $\Psi =\bar{ \psi}\psi $; $g$ is the determinant of the
metric tensor $g_{\mu \nu }$; R is the Ricci scalar; $\psi $ is two
components, particle and anti-particle, Dirac field; $\bar{\psi}$ is
adjoint of the $\psi $ and $\bar{\psi}=\psi ^{\dag }\sigma ^{3}$. In
this action, $\Omega _{\mu }(x)$ are spin connection and are given
as
\begin{equation}
\Omega _{\mu }(x)=\frac{1}{4}g_{\lambda \alpha }(e_{\nu ,\mu
}^{i}e_{i}^{\alpha }-\Gamma _{\nu \mu }^{\alpha })s^{\lambda \nu
}(x), \label{spin}
\end{equation}%
where $\Gamma _{\nu \mu }^{\alpha }$ is Christoffell symbol, and
$g_{\mu \nu }$ is given in term of triads, $e_{\mu }^{(i)}(x),$ as
follows,
\begin{equation}
g_{\mu \nu }(x)=e_{\mu }^{i}(x)e_{\nu }^{j}(x)\eta _{ij},
\label{met2}
\end{equation}%
where $\mu $ and $\nu $ are curved spacetime indices running from
$0$ to $2$. $i$ and $j$ are flat spacetime indices running from $0$
to $2$ and $\eta_{ij}$ is the $2+1$ dimensional Minkowskian metric
with signature (1,-1,-1). The $s^{\lambda \nu }(x),$ spin operators,
are given by
\begin{equation}
s^{\lambda \nu }(x)=\frac{1}{2}[\overline{\sigma }^{\lambda
}(x),\overline{ \sigma }^{\nu }(x)],  \label{met2}
\end{equation}%
where $\bar{\sigma}^{\mu }(x)$ are the spacetime dependent Dirac
matrices in the $2+1$ dimensional. Thanks to triads, $e_{(i)}^{\mu
}(x)$, $\bar{\sigma}^{\mu }(x)$ are related to the flat spacetime
Dirac matrices, $\overline{\sigma }^{i}$, as follows
\begin{equation}
\bar{\sigma}^{\mu }(x)=e_{(i)}^{\mu }(x)\bar{\sigma}^{i},
\label{met3}
\end{equation}%
where $\bar{\sigma}^{i}$ are
\begin{equation}
\bar{\sigma}^{0}=\sigma ^{3}\ \ ,\bar{\sigma}^{1}=i\sigma ^{1},\ \bar{\sigma}%
^{2}=i\sigma ^{2}.  \label{met4}
\end{equation}
$\sigma ^{1}$, $\sigma ^{2}$ \ and $\sigma ^{3}$ are Pauli matrices
\cite {sucu}. In this representation, the Dirac equation gives an
important information about the curved spacetime
\cite{pit,unver,gecim}. To analyse the expansion of the universe, we
will consider the spatially flat spacetime background in $2+1$
dimensional which is analogous of the 3+1 dimensional
Friedmann-Robertson-Walker metric as follows,

\begin{equation}  \label{FRW}
ds^2 = dt^2 - a^2(t)[ dx^2 + dy^2],
\end{equation}
where $a(t)$ is the scale factor of the Universe. The scalar
curvature corresponding to the FRW metric (\ref{FRW}) takes the form $R=-2 \left(\frac{%
2\ddot{a}}{a}+\frac{\dot{a}^{2}}{a^{2}}\right)$, where the dot
represents differentiation with respect to cosmic time $t$. Given
the background in Eq.(\ref{FRW}), it is possible to obtain the
point-like Lagrangian from action (\ref{act1}) in the following form
\begin{eqnarray}  \label{lag}
L = 2 F \dot{a}^{2} + 4 F^{\prime }a \dot{a}\dot{\Psi} + \frac{\imath a^2}{2}\left(%
\bar{\psi}\sigma^{3}\dot{\psi}-\dot{\bar{\psi}}\sigma^{3}\psi\right)
-a^{2} V.
\end{eqnarray}
Here the prime denotes the derivative with respect to the bilinear
$\Psi$. Because of homogeneity and isotropy of the metric, it is
assumed that the spinor field only depends on time $t$, i.e. $\psi =
\psi (t)$. The Dirac's equations for the spinor field $\psi$ and its
adjoint $\bar{\psi}$ are obtained from the point-like Lagrangian
(\ref{lag}) such that the Euler-Lagrange equations for $\psi$ and
$\bar{\psi}$ are
\begin{eqnarray}
& & \dot{\psi} + H \psi + i V^{\prime}\sigma^{3}\psi +2 i F^{\prime
}(2\dot{H}+3
H^2)\sigma^{3}\psi = 0,  \label{dirac1} \\
& & \dot{\bar{\psi}} + H \bar{\psi} - i
V^{\prime}\bar{\psi}\sigma^{3}-2 i F^{\prime }(2\dot{H}+3
H^2)\bar{\psi} \sigma^{3}=0,  \label{dirac2}
\end{eqnarray}
where $H = \dot{a}/a$ denotes the Hubble parameter. On the other
hand, from the point-like Lagrangian (\ref{lag}) and by considering
the Dirac's equations, we find the second order Euler-Lagrange
equation for $a$, i.e. the acceleration equation,
\begin{equation}  \label{acce}
\frac{\ddot{a}}{a} = -\frac{p_{_f}}{2 F}.
\end{equation}
Finally, we also consider the Hamiltonian constraint equation ($E_L
= 0$) associated with the Lagrangian (\ref{lag})
\begin{equation}  \label{hamilton}
E_{L} = \frac{\partial {L}}{\partial{\dot{a}}} \dot{a} + \frac{\partial {L}}{%
\partial{\dot{\psi}}} \dot{\psi} +{\dot{\bar{\psi}}} \frac{\partial {L}}{%
\partial{\dot{\bar{\psi}}}} -L ,
\end{equation}
which yields the Friedmann equation as follows
\begin{equation}  \label{fried}
H^2 = \frac{\rho_{_f}}{2 F}.
\end{equation}
In Eqs. (\ref{acce}) and (\ref{fried}), $\rho_{_f}$ and $p_{_f}$ are
the effective energy density and pressure of the fermion field,
respectively, so that they have the following form
\begin{equation}  \label{ener}
\rho_{_f} = -(4 F^{\prime }H \dot{\Psi} + V) ,
\end{equation}
\begin{eqnarray}  \label{pressure}
p_{_f} = 2 F^{\prime }(\ddot{\Psi} + H \dot{\Psi})+ 2 F^{\prime
\prime }\dot{\Psi}^{2}\nonumber \\
- \left[2 F^{\prime }(2 \dot{H} + 3 H^2 ) + V^{\prime }\right]\Psi + V%
.
\end{eqnarray}
In order to solve the field equations, we have to choose a form for
the coupling function and for the potential density. To do this, in
the following section we will use the Noether gauge symmetry
approach.

\section{The Noether gauge symmetry approach} \label{ns}

Thanks to Pauli matrices, in terms of the components of the spinor field $%
\psi= (\psi_1, \psi_2)^T$ and its adjoint $\bar{\psi} = ({\psi_1}^\dagger, -{%
\psi_2}^\dagger )$, the Lagrangian (\ref{lag}) can be rewritten as
follows
\begin{eqnarray}  \label{la2}
L = 2 F \dot{a}^{2} + 4 F^{\prime }a \dot{a}\sum_{i=1}^2 \epsilon_{i}(\dot{%
\psi_{i}^\dagger}\psi_{i} + \psi_{i}^\dagger\dot{\psi_{i}})\nonumber \\
+ \frac{i a^2}{2}\left[\sum_{i}^2(\psi_{i}^\dagger\dot{\psi_{i}} - \dot{%
\psi_{i}^\dagger}\psi_{i})\right] -a^{2} V,
\end{eqnarray}
where $\epsilon_{i}=\left\{
\begin{array}{c}
1~~ ~~\textstyle{for}~~ i=1\;\cr -1~~ \textstyle{for}~~ i=2\;%
\end{array}
\right.. $

Noether theorem is useful tool in theoretical physics which states
that any differentiable symmetry of an action of a physical system
leads to a corresponding conserved quantity \cite{noo}. The idea to
use Noether symmetry approach without gauge term in generalized
theories of gravity studies is not new and was first introduced by
Cappoziello et al. to find new cosmological solutions. An another
technique is related with the more general symmetries known as
Noether gauge symmetries  which include non-zero gauge term
\cite{jamil,hussain11}. Taking into account a gauge term in Noether
symmetry equation gives a more general definition of the Noether
symmetry.

A vector field $\mathbf{X}$ for the point Lagrangian (\ref{la2}) is
\begin{eqnarray}
\mathbf{X} =\tau \frac{\partial}{\partial t}+\alpha
\frac{\partial}{\partial a}
+ \sum_{i=1}^{2}\left(\beta_{i} \frac{\partial}{\partial \psi_{i}} +  \gamma_{i} \frac{%
\partial}{\partial \psi_{i}^\dagger} \right) ,  \label{vecf}
\end{eqnarray}
where $\alpha, \beta_{j}$ and $\gamma_{j}$ are depend on $t, a, \psi_{j}$ and $%
\psi_{j}^\dagger$ and they are determined from the Noether gauge
symmetry condition. The first prolongation of $\mathbf{X}$ is given
by
\begin{equation}\label{fpro}
{\bf X^{[1]}} = {\bf X}+\alpha_{t} \frac{\partial}{\partial \dot{a}}
+ \sum_{i=1}^{2} \left(\beta_{i t} \frac{\partial}{\partial
\dot{\psi_{i}}}+\gamma_{i t} \frac{\partial}{\partial
\dot{\psi_{i}^\dagger}} \right) ,
\end{equation}
in which
\begin{eqnarray}\label{defi}
{\alpha_{t}} &=& D_{t}\alpha-\dot{a}D_{t}\tau,  \nonumber\\
{\beta_{jt}} &=& D_{t}\beta_{j}-\dot{\psi_{j}}D_{t}\tau,  \nonumber\\
{\gamma_{jt}} &=& D_{t}\gamma_{j}-\dot{\psi_{j}^\dagger} D_{t}\tau.
\end{eqnarray}
The vector field $\mathbf{X}$ is a Noether gauge symmetry
corresponding to the Lagrangian $L(t,a,\psi_{j}, \psi_{j}^\dagger,
\dot{a},\dot{\psi_{j}}, \dot{\psi_{j}^\dagger})$, if the condition
\begin{equation}\label{noether}
{\bf X}^{[1]}L + LD_{t}(\tau) = D_{t}B,
\end{equation}
holds, where $B(t,a,\psi_{j}, \psi_{j}^\dagger,
\dot{a},\dot{\psi_{j}}, \dot{\psi_{j}^\dagger})$ is a gauge function
and $D_{t}$ is the operator of total differentiation with respect to
$t$
\begin{equation}\label{totd}
{D_{t}} =  \frac{\partial}{\partial t} + \dot{a}
\frac{\partial}{\partial a} + \sum_{i=1}^{2}\left(\dot{\psi_{i}}
\frac{\partial}{\partial \psi_{i}} +  \dot{\psi_{i}^\dagger}
\frac{\partial}{\partial \psi_{i}^\dagger}\right).
\end{equation}
The significance of Noether gauge symmetry is clearly comes from the
fact that if the vector field $\mathbf{X}$ is the Noether gauge
symmetry corresponding to the Lagrangian $L(t,a,\psi_{j},
\psi_{j}^\dagger, \dot{a},\dot{\psi_{j}}, \dot{\psi_{j}^\dagger})$,
then
\begin{eqnarray}\label{Frsti}
 I = \tau L + \left(\alpha-\tau \dot{a}\right) \frac{\partial
L}{\partial \dot{a}}- B \nonumber \\
+ \sum_{j=1}^{2}\left( (\beta_{j}-\tau
\dot{\psi_{j}}) \frac{\partial L}{\partial \dot{\psi_{j}}} +
(\gamma_{j}-\tau \dot{\psi_{j}^\dagger}) \frac{\partial L}{\partial
\dot{\psi_{j}^\dagger}}\right)
\end{eqnarray}
is a first integral or a conserved quantity associated with
$\mathbf{X}$. Hence the Noether gauge symmetry condition
(\ref{noether}) for the Lagrangian (\ref{la2}) leads to the
following the over-determined system of differential equations
\begin{eqnarray}
2 F \frac{\partial \alpha}{\partial a} + F^{\prime }\sum_{i=1}^2
\epsilon_{i}\left(\beta_{i}\psi_{i}^\dagger +
\gamma_{i}\psi_{i}\right) - F \frac{\partial \tau}{\partial t} \nonumber \\
+ 2 F^{\prime }a \sum_{i=1}^2 \epsilon_{i}
\left(\frac{\partial{\beta_{i}}}{
\partial{a}}\psi_{i}^\dagger + \frac{\partial{\gamma_{i}}}{\partial{a}}
\psi_{i}\right) = 0,  \label{neq1}
\end{eqnarray}
\begin{eqnarray}
F^{\prime }\frac{\partial{\alpha}}{\partial{\psi_{j}}} =0,\quad\quad
F^{\prime }\frac{\partial{\alpha}}{\partial{\psi_{j}^\dagger}} =0,
\label{neq2}
\end{eqnarray}
\begin{eqnarray}
F^{\prime }\frac{\partial{\tau}}{\partial{\psi_{j}}} =0,\quad\quad
F^{\prime }\frac{\partial{\tau}}{\partial{\psi_{j}^\dagger}} =0,
\qquad \frac{\partial{\tau}}{\partial{a}} =0 \label{neq22}
\end{eqnarray}
\begin{eqnarray}
\imath \alpha \psi_{j} + \frac{\imath a}{2} \beta_{j} - \frac{\imath
a}{2} \sum_{i=1}^2\left(\frac{\partial{\beta_{i}
}}{\partial{\psi_{j}^\dagger}}\psi_{i}^\dagger - \frac{\partial{\gamma_{i}}}{%
\partial{\psi_{j}^\dagger}}\psi_{i}\right) \nonumber\\
- 4F'\epsilon_{j}\psi_{j}\frac{\partial{\alpha}}{\partial{t}} + a V
\frac{\partial{\tau}}{\partial{\psi_{j}^\dagger}} +
\frac{1}{a}\frac{\partial{B}}{\partial{\psi_{j}^\dagger}} = 0,
\label{neq3}
\end{eqnarray}
\begin{eqnarray}
\imath \alpha \psi_{j}^\dagger + \frac{\imath a}{2} \gamma_{j} +
\frac{\imath a}{2} \sum_{i=1}^2\left(\frac{\partial{\beta_{i}
}}{\partial{\psi_{j}}}\psi_{i}^\dagger -
\frac{\partial{\gamma_{i}}}{
\partial{\psi_{j}}}\psi_{i}\right)\nonumber\\
+4F'\epsilon_{j}\psi_{j}^\dagger\frac{\partial{\alpha}}{\partial{t}}
- a V \frac{\partial{\tau}}{\partial{\psi_{j}}} -
\frac{1}{a}\frac{\partial{B}}{\partial{\psi_{j}}}=0,  \label{neq4}
\end{eqnarray}
\begin{eqnarray}
F^{\prime }\epsilon_{j}\psi_{j}^\dagger\left(\alpha + a
\frac{\partial{\alpha}}{\partial{ a}}\right) +F
\frac{\partial{\alpha}}{\partial{\psi_{j}}} \nonumber\\
+F^{\prime }a \left[\epsilon_{j} \gamma_{j}-\epsilon_{j}\psi_{j}^\dagger \frac{\partial {\tau}}{\partial{t}} +\sum_{i=1}^2\epsilon_{i}\left(%
\frac{\partial{\beta_{i}}}{\partial{\psi_{j}}}\psi_{i}^\dagger +\frac{%
\partial{\gamma_{i}}}{\partial{\psi_{j}}}\psi_{i}\right)\right]  \nonumber \\
+F^{\prime \prime } \epsilon_{j} a \psi_{j}^\dagger \sum_{i=1}^2
\epsilon_{i}(\beta_{i}\psi_{i}^\dagger + \gamma_{i}\psi_{i})= 0,
\label{neq5}
\end{eqnarray}
\begin{eqnarray}
F^{\prime }\epsilon_{j}\psi_{j}\left(\alpha + a
\frac{\partial{\alpha}}{\partial{ a}}\right) +F
\frac{\partial{\alpha}}{\partial{\psi_{j}^\dagger}}\nonumber\\
+F^{\prime }a \left[\epsilon_{j} \beta_{j}-\epsilon_{j}\psi_{j}\frac{\partial {\tau}}{\partial{t}} +\sum_{i=1}^2\epsilon_{i}\left(%
\frac{\partial{\beta_{i}}}{\partial{\psi_{j}^\dagger}}\psi_{i}^\dagger +\frac{%
\partial{\gamma_{i}}}{\partial{\psi_{j}^\dagger}}\psi_{i}\right)\right]  \nonumber \\
+F^{\prime \prime } \epsilon_{j} a \psi_{j} \sum_{i=1}^2
\epsilon_{i}(\beta_{i}\psi_{i}^\dagger + \gamma_{i}\psi_{i})= 0,
\label{neq6}
\end{eqnarray}
\begin{eqnarray}
4F \frac{\partial {\alpha}}{\partial {t}}+\frac{\imath
a^2}{2}\sum_{i=1}^2
\left(\frac{\partial{\beta_{i}}}{\partial{a}}\psi_{i}^\dagger -
\frac{\partial{\gamma_{i}}}{\partial{a}}\psi_{i}\right)\nonumber\\
- a^2 V
\frac{\partial {\tau}}{\partial {a}}-\frac{\partial {B}}{\partial
{a}} + 4F'a\sum_{i=1}^2
\epsilon_{i}\left(\frac{\partial{\beta_{i}}}{\partial{t}}\psi_{i}^\dagger
+ \frac{\partial{\gamma_{i}}}{\partial{t}}\psi_{i}\right) =0, \label{neq7}
\end{eqnarray}

\begin{eqnarray}
(2 \alpha+a \frac{\partial{\tau}}{\partial {t}}) V
+\frac{1}{a}\frac{\partial{B}}{\partial {t}} + a V^{\prime
}\sum_{i=1}^2 \epsilon_{i} \left(\beta_{i}\psi_{i}^\dagger +
\gamma_{j}\psi_{i}\right) \nonumber\\
-\frac{\imath
a}{2}\sum_{i=1}^2\left(\psi_{i}^\dagger
\frac{\partial{\beta_{i}}}{\partial{t}}-\psi_{i}
\frac{\partial{\gamma_{i}}}{\partial{t}}\right) = 0. \label{neq9}
\end{eqnarray}
This system given by Eqs. (\ref{neq1})-(\ref{neq9}) are obtained by
imposing
the fact that the coefficients of $\dot{a}^2,\dot{a}, \dot{\psi_{j}}, \dot{%
\psi_{j}^\dagger}, \dot{a}\dot{\psi_{j}}$,
$\dot{\psi_{j}}\dot{\psi_{l}}$  and so on, vanish. Now we will
search a solution of Eqs. (\ref{neq1})-(\ref{neq9}). The equations
(\ref{neq2})
give us two cases: $F^{\prime }=0$ or $\frac{\partial{\alpha}}{\partial{\psi_{l}%
}}=0$ and $\frac{\partial{\alpha}}{\partial{\psi_{l}^\dagger}}=0$ ($%
F^{\prime }\neq0$). In this study, we will neglect the case
$F^{\prime }=0$ that corresponds to the fermionic field minimal
coupled to gravity. From the equations (\ref{neq22}), one can
immediately sees $\tau=\tau(t)$. From the rest Noether gauge
symmetry equations, the complete solution is obtained as follows
\begin{eqnarray}\label{vec1}
\alpha &=& -\frac{c_{1}}{2(n-1)} a,  \nonumber\\
\beta_{j} &=& \frac{c_{1}}{2(n-1)}\psi_{j}-\epsilon_{j} c_{2}\psi_{j},  \nonumber\\
\gamma_{j} &=& \frac{c_{1}}{2(n-1)}\psi_{j}^{\dagger}+\epsilon_{j} c_{2}\psi_{j}^{\dagger}, \nonumber\\
\tau &=& c_{1}t+c_{3}, \quad B=c_{4},
\end{eqnarray}
and the coupling and the potential function are power law forms of
the function of the bilinear $\Psi$, i.e.,
\begin{eqnarray}
F(\Psi) = f_{0}\Psi^{n},  \label{coupling}
\end{eqnarray}
\begin{eqnarray}
V(\Psi) = \lambda\Psi^{2-n},  \label{pot1}
\end{eqnarray}
where $c_l$, $\lambda$, $f_{0}$ and $n$ ($n\neq 1$) are integration
constants.

From the vector field (\ref{vec1}), the Lagrangian (\ref{la2})
admits three Noether gauge symmetries which are
\begin{eqnarray}
{\bf X}_{1} = \frac{\partial}{\partial{t}},  \label{gen1}
\end{eqnarray}
\begin{eqnarray}
{\bf X}_{2} = t\frac{\partial}{\partial{t}} -
\frac{1}{2(n-1)}\left[a\frac{\partial}{\partial{a}}-\sum_{i=1}^2(\psi_{i}\frac{\partial}{\partial{\psi_{i}}}+
\psi_{i}^{\dagger}\frac{\partial}{\partial{\psi_{i}}^{\dagger}})\right],
\label{gen2}
\end{eqnarray}
\begin{eqnarray}
{\bf X}_{3} =
-\sum_{i=1}^2\epsilon_{i}(\psi_{i}\frac{\partial}{\partial{\psi_{i}}}-\psi_{i}^{\dagger}\frac{\partial}{\partial{\psi_{i}}^{\dagger}}).
\label{gen3}
\end{eqnarray}
These generators constitute a well-known three-dimensional Lie
algebra with commutation relations
\begin{eqnarray} \label{commu1}
[{\bf X}_{1},{\bf X}_{2}] =  {\bf X}_{1}, \quad [{\bf X}_{1},{\bf
X}_{3}] =[{\bf X}_{2},{\bf X}_{3}] =0
\end{eqnarray}
The three first integrals (or conserved quantities) associated with
the Noether gauge symmetries are
\begin{eqnarray} \label{frstI-1}
I_{1} = -2 F'\dot{a}^2-4 F'a \dot{a}\dot{\Psi}-a^2 V,
\end{eqnarray}
\begin{eqnarray} \label{frstI-2}
I_{2} =
tI_{1}-\frac{2a}{n-1}\left(F\dot{a}-2F'\Psi\dot{a}+F'a\dot{\Psi}\right)
,
\end{eqnarray}
\begin{eqnarray} \label{frstI-3}
I_{3} = \imath a^2 \Psi.
\end{eqnarray}
Here the constant parameter $c_{4}$ is assumed to be zero in the
gauge function $B$. We note that the first integral (\ref{frstI-1})
is related with the energy function (\ref{hamilton}), so that the
first integral $I_{1}$ vanishes.

\section{The solutions of field equations} \label{fieldeqns}
Since the coupling function $F$ depends on the bilinear function
$\Psi$, from Dirac's equations (\ref{dirac1}) and (\ref{dirac2}) one
gets
\begin{eqnarray}
\dot{\Psi} + 2 \frac{\dot{a}}{a}\Psi=0,  \label{sol}
\end{eqnarray}
which integrates to give
\begin{eqnarray}
\Psi = \frac{\Psi_0}{a^2},  \label{sol2}
\end{eqnarray}
where $\Psi_0$ is a constant of integration. Inserting the equation
(\ref{sol2}) into the first integral equation (\ref{frstI-3}), we
get $I_{3}=\imath \Psi_{0}$. Considering the equations (\ref{sol2})
and the coupling function (\ref{coupling}), the first integral
(\ref{frstI-2}) can be rewritten as
\begin{eqnarray}
\dot{a}-k a^{2n-1} = 0,  \label{frstI-22}
\end{eqnarray}
where we define $k=\frac{(n-1)I_{2}}{2(4n-1)f_{0}\Psi_{0}^n}$, and
$n\neq 1/4$. Now the the equation (\ref{frstI-22}) can be used to
find out the time dependence of the cosmic scale factor as follows
\begin{eqnarray}
a(t) = \left[2k(1-n) t +a_{0}\right]^{\frac{1}{2(1-n)}},
\label{scale1}
\end{eqnarray}
where $a_{0}$ is an integration constant. Using the solution
(\ref{scale1}) in the Friedmann equation (\ref{fried}) with
(\ref{ener}) and the acceleration equation (\ref{acce}) with
(\ref{pressure}), we obtain a constraint relations between the
constants as $\lambda=\frac{I_{2}^2
(n-1)^2}{2\Psi_{0}^2f_{0}(4n-1)}$.   Therefore, the solution for the
cosmic scale factor that is obtained from the Noether gauge symmetry
represents a power law expansion for the universe. It is remarkable
from (\ref{scale1}) that models with model parameter $n>1/2$ obey an
accelerated power law expansion while for $n < 1/2$ a decelerated
expansion occurs. For $n=1/2$, we have
\begin{eqnarray}
a(t)=-\frac{I_{2}}{4f_{0}\sqrt{\Psi_{0}}}t + a_0. \label{scl2}
\end{eqnarray}
This solution correspond to the matter dominant universe with the
pressure of the Dirac field $p_{_f}=0$  in the General Relativity in
$2+1$ dimensions. Therefore, this solution shows that the Dirac
field behaves as a standard pressureless matter field in the $2+1$
dimensions.

For the our model, we search that whether the fermionic field can
provide alternative for dark energy or not. For this purpose we can
define the equation of state parameter of the fermionic field by
using the energy density (\ref{ener}) and pressure (\ref{pressure})
as $\omega_{f}=p_{_f}/\rho_{_f}$. Considering the equations
(\ref{coupling}), (\ref{pot1}), (\ref{sol2}) and (\ref{scale1}), we
obtain
\begin{eqnarray}
\omega_{_f} = 1-2n.  \label{eos}
\end{eqnarray}
According to astrophysical data, the equation of state parameter
tend to value $-1$. For the equation of state parameter less than
$-1$ the dark energy is described by phantom, for $-1 < \omega <
-1/3$ the quintessence dark energy is observed and the case $\omega
=-1$ corresponds to the cosmological constant. For the our model, if
$2/3 <n < 1$, we obtain the quintessence phase, if $n > 1$, we have
the phantom phase. In both cases, the universe is both expanding and
accelerating. Therefore, the results show that the fermionic field
may behave like both quintessence and phantom dark energy field in
the late-time universe.

Now, we return the case $n=1$ so that the coupling and potential
functions have linear forms of $\Psi$ from the equation
(\ref{coupling}) and (\ref{pot1}) as follows
\begin{eqnarray}
F(\Psi) = f_{0}\Psi,  \label{coupling2}
\end{eqnarray}
\begin{eqnarray}
V(\Psi) = \lambda\Psi.  \label{pot2}
\end{eqnarray}
For this case, we do not need to the first integral obtained by
Noether gauge symmetry approach because the Friedman equation
(\ref{fried}) and acceleration equation (\ref{acce}) can be directly
integrated by usually method. Friedmann equation (\ref{fried}) is
reduced to
\begin{eqnarray}
\frac{\dot{a}}{a} -\sqrt{\frac{\lambda}{6 f_{0}}} =0,  \label{sol-4}
\end{eqnarray}
which has the solution
\begin{eqnarray}
a(t) =a_0 e^{H_0 t}, \quad \mathrm{where} \quad
H_0=\sqrt{\frac{\lambda}{6 f_{0}}} \label{sol-5}
\end{eqnarray}
and $a_0$ is a constant. It is clear that this solution describes an
inflationary period, where the cosmic scale factor increases
exponentially with the cosmic time. Therefore, one can say that the
Dirac field can be behaved as inflaton field in the $2+1$
dimensions. From Eqs. (\ref{ener}) and (\ref{pressure}), the energy
density and the pressure of the Dirac field are given by
\begin{eqnarray}
\rho_{_f} = \frac{\lambda \Psi_0}{3 a_0^2}e^{-2 H_0 t}, \qquad
p_{_f}=-\rho_{_f}, \label{sol-eden}
\end{eqnarray}
with the equation of state parameter $\omega_{_f}= -1$.

\section{Concluding remarks}  \label{conc}

The study of cosmologically models in the $2+1$ dimensional
gravitation theories provide mathematical simplicity in
understanding the physical models. In the present study, we have
considered a theory including the Dirac field which is non-minimally
coupled to the gravity in $2+1$ dimension. Using the Noether gauge
symmetry approach, we have determined the explicit forms of the
coupling function and potential as a power-law functions of the
bilinear $\Psi$ given by (\ref{coupling}) and (\ref{pot1}),
respectively. The solutions of the field equations for FRW spacetime
are presented by using the results obtained from the Noether gauge
symmetry approach. It is shown that in the general case $n$, the
Dirac field play role of the dark energy in the late-time universe.
For the special case $n=1/2$, the solution of dynamical equation
describes a decelerated universe with a matter dominated behavior in
$2+1$ dimension. We also consider a model where the coupling
function and the potential have linear forms of $\Psi$ (i.e. the
case $n=1$). For such a model, the cosmological solution describes
an inflationary period for the early-time universe. Therefore, we
may conclude that the Dirac field behaves as an inflation field.

\begin{acknowledgments}
This work was supported by the Scientific Research
Projects Unit of Akdeniz University.
\end{acknowledgments}
\newpage

\begin{thebibliography}{99}


\bibitem{ade1} P.A.R. Ade et al., Phys. Rev. Lett. 112, 241101 (2014); P.A.R. Ade et al., Astron. Astrophys. 571, A1 (2014) \emph{Planck Collaboration}, arXiv:1303.5076.

\bibitem{sper1} D.N. Spergel et al., Astrophys. J. Suppl. 148, 175 (2003).

\bibitem{sper2} D.N. Spergel et al., Astrophys. J. Suppl. 170, 377 (2007).

\bibitem{alan} A.H. Guth, Phys. Rev. D 23, 347 (1981).

\bibitem{linde} A.D. Linde, Phys. Lett. B 129,177 (1983).

\bibitem{riess99}  A.G. Riess et al., Astrophys. J. 116, 1009 (1998).

\bibitem{Perll} S. Perlmutter et al., Astrophys. J. 517, 565 (1999).

\bibitem{netterfield02} C.B. Netterfield et al., Astrophys. J. 571, 664 (2002).

\bibitem{bennet} C.N. Bennett et al., Astrophys. J. Suppl. 148, 1 (2003).

\bibitem{tegmark04} M. Tegmark et al., Phys. Rev. D 69, 103501 (2004).

\bibitem{Rat} B. Ratra and P.J.E. Peebles, Phys. Rev. D 37, 3406 (1988).

\bibitem{Cad} R.R. Caldwell, Phys. Lett. B 545, 23 (2002).

\bibitem{Sot} T.P. Sotiriou and V. Faraoni, Rev. Mod. Phys. 82, 451 (2010); A. De Felice and S. Tsujikawa, Living Rev. Relativ. 13, 3 (2010); S. Nojiri and S.D. Odintsov, Phys. Rep. 505, 59 (2011);  S. Capozziello and M. de Laurentis, Phys. Rep. 509, 167 (2011).

\bibitem{Beng} G.R. Bengochea and R. Ferraro, Phys. Rev. D 79, 124019 (2009); E.V. Linder, Phys. Rev. D 81, 127301 (2010); P. Wu and H.W. Yu, Phys. Lett. B 693, 415 (2010); R. Myrzakulov, Eur. Phys. J. C 71, 1752 (2011).

\bibitem{Saha1} B. Saha, Phys. Rev. D 64, 123501 (2001).

\bibitem{Ar} C. Armendariz-Picon and P.B. Greene, Gen. Relativ. Gravit. 35, 1637 (2003).

\bibitem{Saha2} B. Saha and T. Boyadjiev, Phys. Rev. D 69, 124010 (2004).

\bibitem{Rib1} M.O. Ribas, F.P. Devecchi and G.M Kremer, Phys. Rev. D 72, 123502 (2005).

\bibitem{Rib2} M.O. Ribas, F.P. Devecchi and G.M Kremer, Europhys Lett. 81, 19001 (2008).

\bibitem{vakili} B. Vakili, S. Jalalzadeh and H.G. Sepangi, J. Cosmol. Astropart. Phys. 05, 006 (2008).

\bibitem{Kremer2} L.L. Samojeden, F.P. Devecchi and G.M. Kremer, Phys. Rev. D 81, 027301 (2010).

\bibitem{Kremer} R.C. De Souza and G.M. Kremer, Class. Quantum Grav. 25, 225006 (2008).

\bibitem{capo1} S. Capozziello and R. de Ritis, Phys. Lett. A 177, 1 (1993).

\bibitem{capo2} S. Capozziello, R. de Ritis, C. Rubano and P. Scudellaro, Riv. Nuovo Cimento 19, 1 (1996); U. Camci and Y. Kucukakca, Phys. Rev. D 76, 084023 (2007); Y. Kucukakca, U. Camci and I. Semiz, Gen. Relativ. Gravit. 44, 1893 (2012); A. Paliathanasis, M. Tsamparlis, S. Basilakos and S. Capozziello, Phys. Rev. D 89, 063532 (2014).

\bibitem{cap03} S. Capozziello and A. de Felice, J. Cosmol. Astropart. Phys. 08, 016 (2008); B. Vakili, Phys. Lett. B 664, 16 (2008); M. Roshan and F. Shojai, Phys. Lett. B 668, 238 (2008); Y. Kucukakca and U. Camci, Astrophys. Space Sci. 338, 211 (2012).

\bibitem{wei} H. Wei , X-J. Guo and L.F. Wang, Phys. Lett. B 707, 298 (2012); A. Paliathanasis, S. Basilakos, E.N. Saridakis, S. Capozziello, K. Atazadeh, F. Darabi and M. Tsamparlis, Phys. Rev. D 89, 104042 (2014).

\bibitem{kuc} Y. Kucukakca, Eur. Phys. J. C 73, 2327 (2013).

\bibitem{wit} E. Witten, Nuc. Phys. B 311, 46 (1998).

\bibitem{deser} S. Deser, R. Jackiw R and G. 't Hooft, Annals of Phys. 152, 220-235 (1984).

\bibitem{cor} N.J. Cornish and N.E. Frankel, Phys. Rev. D 43, 2555 (1991).

\bibitem{banados} M. Banados, C. Teitelboim and J. Zanelli, Phys. Rev. Lett. 69, 1849 (1992).

\bibitem{sucu} Y. Sucu and N. Unal, J. Math. Phys. 48, 052503 (2007).

\bibitem{pit} J.P.M. Pitelli and P.S. Letelier, Phys. Rev. D 77, 124030 (2008).

\bibitem{unver} O. Unver and O. Gurtug, Phys. Rev. D 82, 084016 (2010).

\bibitem{gecim} G. Gecim and Y. Sucu, J. Cosmol. Astropart. Phys. 02, 023 (2013).

\bibitem{noo} E. Noether, Gott.Nachr. 235-257, (1918) ; English translationn in \textit{Transport Theory and
Statistical Physics,} \textbf{1} 186-207, arXiv:physics/0503066v1.

\bibitem{jamil} M. Jamil, F.M. Mahomed and D. Momeni, Phys. Lett. B 702, 315 (2011).

\bibitem{hussain11} I. Hussain, M. Jamil and F.M. Mahomed, Astrophys. Space Sci. 337, 373 (2012).


\end{thebibliography}

\end{document}